\documentclass{emulateapj}
\usepackage[usenames]{color}
\usepackage{apjfonts,amsmath}
\usepackage{color}
\usepackage[english]{babel}
\usepackage{amssymb}
\shorttitle{ULBDM and CMB}
\shortauthors{Rodr\'{\i}guez-Montoya, Maga\~na, Matos \& P\'erez-Lorenzana}

\begin{document}
\title{Ultra light bosonic dark matter and cosmic microwave background}

\author{Iv\'an Rodr\'{\i}guez-Montoya\altaffilmark{1,3}, Juan Maga\~na\altaffilmark{2,3},
Tonatiuh Matos\altaffilmark{1,3} and Abdel P\'erez-Lorenzana\altaffilmark{1,3}}
\affil{$^1$Departamento de F{\'\i}sica, Centro de
Investigaci\'on y de Estudios Avanzados del IPN, A.P. 14-740,
07000 Mexico City, Mexico; \\
rrodriguez@fis.cinvestav.mx, tmatos@fis.cinvestav.mx, aplorenz@fis.cinvestav.mx\\
$^2$Instituto de Astronom\'{\i}a, Universidad Nacional Aut\'onoma de M\'{e}xico, 
Ciudad Universitaria, 04510 Mexico City, Mexico; jmagana@astroscu.unam.mx}

\altaffiltext{3}{Part of the Instituto Avanzado de Cosmolog\'ia (IAC) collaboration http://www.iac.edu.mx/}

\begin{abstract}

In this paper, we consider the hypothesis in which a species of ultra light bosonic dark matter (ULBDM)
with mass $m_{B}\sim 10^{-22}$ eV could be the dominant dark matter (DM) in the Universe.
As a first approach we work in the context of kinetic theory,
where ULBDM is described by the phase space distribution function
whose dynamics is dictated by the Boltzmann-Einstein equations.
We investigate the effects that this kind of dark matter imprints in the acoustic peaks of the cosmic microwave background.
We find that the effect of the Bose-Einstein statistics is small, albeit perceptible,
and is equivalent to an increase of non-relativistic matter.
It is stressed that in this approach,
the mass-to-temperature ratio necessary for ULBDM to be a plausible DM candidate is about five orders of magnitude.
We show that reionization is also necessary and we address a range of consistent values for this model.
We find that the temperature of ULBDM is below the critical value,
impliying that Bose-Einstein condensation is inherent to the ULBDM paradigm.

\end{abstract}


\section{Introduction}

One of the most precise cosmological observations is the measurement of the anisotropies in the cosmic microwave background (CMB).
The experimental data are useful for probing the dynamics and properties of many theoretical cosmological models.
Nowadays, the most successful model describing the observed profiles of CMB anisotropies
is the so called, cold dark matter with a cosmological constant ($\Lambda$CDM).
Nevertheless, the cold dark matter (CDM) model has some inconsistencies
with the observations on galactic and sub-galactic scales.
For instance, CDM predicts cusp central density profiles of dark halos in low surface brightness (LSB) and dwarf galaxies;
meanwhile the measurements indicate a smooth distribution of matter.
Also, CDM has some discrepancies between the number of predicted satellite galaxies in high-resolution N-body
simulations and observations.
In this sense, the possibility of alternative hypothesis on the nature of dark matter (DM) is open.

In recent years, it has been argued that a real scalar field $\Phi$,
minimally coupled to gravity, could be a plausible candidate for dark matter (DM).
This alternative proposal (or similar ideas) is called scalar field dark matter (SFDM)
\citep{ji, sin, lee, hu, matos_1, matos_2, varun, matos_3, lee2, migue}.
Several previous works have shown that a scalar field is able to reproduce the cosmological evolution of the Universe.
To this end, the scalar field is endowed with a scalar potential $V(\Phi)$ of the form $\cosh (\Phi)$ or $\Phi^{2}$
and obeys an equation of state $\omega_{\Phi} \equiv p_{\Phi}/\rho_{\Phi}$
that varies in time ($-1 \leq \omega_{\Phi} \leq 1$)
\citep[see for example][]{matos_3,magana}. \citet{matos_3} found that SFDM model predicts a supression on
the mass power spectrum for small scales.
Thus, SFDM could help to explain the excess of satellite galaxies.

The SFDM paradigm has also been tested on galactic scales,
showing interesting results.
For instance, \citet{bernal} showed that 
the density profiles for SFDM halos are non cuspy profiles,
in accordance with the observations of LSB galaxies \citep[see also][Matos et al. 2009]{bohmer}.
Moreover, it is noticeable that in the relativistic regime, scalar fields can form gravitationally bounded structures.
These are called boson stars for complex scalar fields \citep[][]{ruffini,lee,paco} and oscillatons for real scalar fields \citep[][]{seidel,luis_1, alcubierre}.
There are also scalar field stable gravitational structures described by the Schr\"odinger-Poisson system \citep{newcol,gravcool,argelia}.
One of the most promising and physically interesting feature of SFDM resides on the hypothesis that it describes cosmological Bose-Einstein condensates (BEC)
\citep[see for example][]{woo,luis_2}.
For that reason it is important to provide a thermodynamic understanding of scalar particles,
putting aside for the moment the classical field description.

In the SFDM model, the mass is constrained by phenomenology to an extremely low value ($\sim 10^{-23}\; $eV).
This ultralight scalar field mass fits the observed amount of substructure \citep[][]{matos_3},
the critical mass of galaxies \citep[][]{alcubierre},
the rotation curves of galaxies \citep[][]{bohmer},
the central density profile of LSB galaxies \citep[][]{bernal},
the evolution of the cosmological densities \citep[][]{magana}, etc.
Furthermore, SFDM forms galaxies earlier as CDM;
thus, if SFDM is correct, we expect to see big galaxies at high redshifts.

If this scalar field could be considered as a system of individual
light bosonic particles (with zero spin)
and, moreover, if there are some of these scalar particles in thermal equilibrium
forming an ideal gas, then they should obey the Bose-Einstein statistics.
 From this perspective, ultralight bosonic dark matter (ULBDM)
seems to have some properties close to those of neutrinos.
 In fact, neutrinos constitute a subdominant component of DM in the Universe.
For this reason, it is interesting to mention some of the most remarkable features of the neutrino cosmology.

At very early times of the Universe,
the neutrinos were in thermal equilibrium with the primeval fireball \citep[see for example][]{dodelson_1}.
Due to its low mass compared with its temperature in this epoch,
they behaved exactly as radiation at the moment of its decoupling.
This means that neutrinos fall on the classification of hot dark matter (HDM)
\footnote{The term hot dark matter (HDM) is applied to particles which behave relativistic at the moment of its decoupling.
However, HDM could be non-relativistic today.}.
After decoupling, neutrinos still keep the relativistic distribution,
while they relax only with the expansion of the universe;
this is called the \textit{freeze-out}.
Thus, the temperature of neutrinos evolves simply as $T_{\nu} \propto a^{-1}$
and eventually could reduce to lower values than the mass.
This epoch is known as the \textit{nonrelativistic transition} (NRT) of the neutrino.
Since this epoch, gravitational attraction is sufficient to contribute to structure formation.
Once decoupled and after electron-positron annihilations,
the temperature of neutrinos remains well determined in terms of the temperature of the photons
as $T_{\nu} = (4/11)^{1/3} T_{\gamma}$;
this fixes the neutrino number density today at $n_{\nu}^{(0)} \sim 100 $ cm$^{-3}$.
It is now clear that if NRT occurs earlier,
then neutrinos can form more bounded structures and vice versa.
However, in most of the typical scenarios, this transition occurs too late,
thus making the neutrino contribution subdominant \citep[see an excellent review in][]{pastor}.

Neutrinos and ULBDM are similar in that they are assumed to be in thermal equilibrium
but with negligible couplings with other types of matter.
The value of the mass also ensures that both decouple when still relativistic
and also that their distribution freezes-out.
They differ, however, in many aspects;
an important intrinsic part of the nature of the neutrino is that it is a fermion
and therefore its density has an upper bound.
Thus, the content of neutrinos $\Omega_{\nu}$ is entirely parameterized by its mass,
i.e., $\Omega_{\nu} \approx m_{\nu} / 51.01$ eV  \citep[see][]{kolb}.
In the case of ULBDM, the mass and abundance are not correlated.
Another difference is that couplings of neutrinos are well known
from weak interactions and their thermal decoupling is predicted at $T^{(d)}_{\nu} \sim 1$ MeV. 
However, in the case of ULBDM the energy scale of interactions are unknown,
we shall then assume that decoupling occurs well before neutrino decoupling
and that the temperature $T_{B}$ is as a free parameter.

In the present work, inspired by the neutrino cosmology and the SFDM model,
we assume that the Universe contains two components of DM,
that is, ULBDM and standard CDM.
We focus on the question of whether the dominant contribution of ULBDM
to matter at present epoch could mimic the effects of CDM on the CMB spectrum of anisotropies.
Of course, there is important previous research in this context.
For example, \cite{amendola} treat an \textit{ultra-light pseudo-Goldstone boson} $\phi$ as part of DM
and use CMB data to constrain the density fraction $\Omega_{\phi}/\Omega_{m}$.
This provides a simple and very useful description of \textit{free streaming} effects.
In the context of that work, if ULBDM is assumed to be thermalized,
the sound speed is $c_{eff} \approx T^{(0)}_{B} a^{-1} /m_{B}$ defined under the condition of NRT, i.e., $a \gtrsim T^{(0)}_{B} /m_{B}$.
Another very interesting approach was done by \cite{ferrer},
who analyzed the effects of couplings between baryons and scalar mediatior particles.
In fact, scalar particles in a BEC serves as a thermal bath
for baryons until a time  close to recombination epoch.
One of the most important consequences of these interactions
is a modification in the expression for the speed of sound,
which in turns shifts the position of the acoustic peaks in the CMB.

We stress that in our treatment, ULBDM has a phase-space description,
prescribed by the relativistic kinetic theory,
i.e., the evolution of ULBDM is dictated by the Boltzmann equation coupled to Einstein equations.
This is a novel approach to the paradigm of scalar DM paradigm.
Concretely, the object of treatment in our scheme is neither a classical nor quantum field,
but rather the phase-space distribution function of an ideal gas of individual noninteracting particles.
The scalar particles are thought to be initially thermalized but decoupled from rest of universe.
Even if \textit{a priori} we do not restrict ourselves
to the case in which all the particles reside in a coherent phase,
it is found that Bose-Einstein condensation has a central role in the model.
The BEC formation is assumed to take place before its decoupling during the radiation epoch.
The motivation to work in this scheme is to explore the contribution to the CMB anisotropies
from possible thermal particles filling different energy states in the ULBDM gas.
This is precisely the reason why the name
ULBDM rather than SFDM ismore descriptive in this approach.

In the following, we consider a flat, homogenous and isotropic universe.
We take as fixed parameters the current temperature of the CMB photons T$_{CMB} = 2.726 \;$K,
the current Hubble's constant H$_{0} = 75.0 \;$km s$^{-1}$ Mpc$^{-1}$ ,
and the current baryon density parameter $\Omega_{bar}=0.04$.
Also, we assume, just for simplicity,
that the dark energy in the universe is a cosmological constant $\Lambda$ with a current
density value $\Omega_{\Lambda}=0.74$.
We choose units in which $c = \hbar = k_{B} = 1$,
then, $1 \,$K $ \equiv 8.617 \times 10^{-5} \; $eV.

This paper is organized as follows.
Section \ref{sec:ULBDM} states the key equations of this calculation.
First, we discuss the Bose-Einstein statistics and some concepts of interest,
followed by a brief description of kinetic theory applied to Cosmology.
The physical implications of ULBDM in the CMB anisotropies spectrum are discussed in section \ref{sec:TestCMB}.
Concluding remarks are given in section \ref{sec:Conclusiones}.

All the analysis was done using the public code CMBFAST \cite{cmbfast}.
The calculated curves were compared to the five year \textit{WMAP} satellite data \citep[][]{hinshaw}. (It is available at http://lambda.gsfc.nasa.gov/product/map/current)

 \section{Ultralight bosons as dark matter}\label{sec:ULBDM}
 \subsection{Bose-Einstein Condensation} \label{subsec:BEC}

As stated in the introduction,
we want to explore the hypothesis of the existence of a kind of DM in the universe
composed by scalar particles with an extremely low mass $m_{B} = 10^{-22}$ eV.
We assume that ULBDM was in local thermodynamic equilibrium (LTE)
with the primeval fireball at least in some very early stage of the universe.
Accordingly, it can be defined a temperature $T_{B}$ of the ULBDM,
and the dynamics of these particles may be described by the Bose-Einstein statistics
with a phase-space distribution function
 \begin{equation}
f_{0}(p)=\frac{g_{s}}{e^{(\sqrt{p^{2}+m_{B}^{2}} - \mu)/T_{B}} - 1}, \label{f0}
\end{equation} 
where $g_{s}$ is the number of relativistic degrees of freedom
($g_{s}=1$ in the case of scalar particles) and $\mu$ is the chemical potential.
One immediately finds the condition $\mu \leq m_{B}$
in order to keep the positive value of the distribution function.
In fact, Bose-Einstein condensation occurs when the value of
the chemical potential approaches the mass of particles.
This phenomenom appears for temperatures below a critical value
named \textit{the critical temperature of condensation} $T_{c}$
and consists in a considerable occupation of the state of minimal energy.
We can calculate the number density $n^{(1)}$ of particles from the relativistic kinetic theory of gases (RKT):
\begin{equation}
 n^{(1)} = \int \frac{d^{3}p}{(2\pi)^{3}} f_{0}(p) = \frac{1}{2 \pi^{2}} \int \frac{(E^{2}-m_{B}^{2})^{1/2}E \; dE}{e^{(E - \mu)/T_{B}} - 1},
\end{equation}
where $E^{2}=p^{2}+m^{2}_{B}$ is the energy of each individual particle.
It is natural to assume that the mass-to-temperature ratio was very small
at the moment of decoupling of ULBDM;
thus, we can solve the integral by taking the ultrarelativistic limit ($m_{B} \ll T_{B}$)
which yields, $n^{(1)} = (\zeta(3)/\pi^{2}) T_{B}^{3}$,
with $\zeta(3)\approx 1.2$, the Riemann function.
The critical temperature in the ultrarelativistic regime is then defined as
\begin{equation}
 T_{c} = \left( \frac{\pi^{2} n_{B}}{ \zeta(3)} \right) ^{1/3}.  \label{tempcrit}
\end{equation}
In this equation, $n_{B}$ is the total number density of particles per unit volume;
for $T_{B} > T_{c}$, $n_{B}$ is just $n^{(1)}$.
Quantum statistical mechanics predicts that the occupation of the lower energy state
rapidly increases when the temperature of the Bose gas falls below $T_{c}$.
In this case the total number density is
\begin{equation}
\begin{array}{ll}
 n_{B} = n_{0} + \frac{\zeta(3)}{\pi^{2}} T^{3}_{B}   ,&    T_{B} < T_{c},
\end{array}
\end{equation}
where $n_{0}$ is the particle number density of the BEC.
We can say that ULBDM falls in the classification of HDM,
in the sense that it behaves as radiation at its decoupling epoch.
After this moment, $f_{0}$ is said to be frozen-out until today.
It means that the particles mantain their relativistic distribution
with a temperature scaling as $T_{B} \propto a^{-1}$.
However, relativistic behaviour does not necessarily prevent the BEC formation, \citep[see for example][]{cerci}.
Moreover, because of the expansion of Universe,
ULBDM cools down and the temperature could be under the value necessary for a NRT.
We want to investigate if this could happen and at times early enough to form large-scale structure.

\subsection{Kinetic theory in expansion}
The free evolution of ULBDM is described by the Vlasov equation,
also called the Liouville or collisionless Boltzmann equation \citep[see][]{bernstein}.
On the other hand, the geometry of the universe is described by the FLRW metric
with scale factor $a(\tau)$, perturbed to first order.
ULBDM particles can just move along geodesics
and the Vlasov equation translates this to a differential equation
for the phase-space distribution function $f$ of the ULBDM gas.
In the conformal newtonian gauge, the perturbed metric reads
\begin{equation}
ds^{2}=a^{2}\left\lbrace -(1+2\psi)d\tau^{2}+(1-2\phi)dx^{i}dx_{i}\right\rbrace,
\end{equation}
where $\tau$ is the proper time, and $\psi$ and $\phi$ are the scalar modes of the perturbation.
In this gauge, the tensor and vector degrees of freedom are eliminated from the begining.
Following the same formalism developed for the fermionic sector \citep[see][]{chunga},
it is useful to define the \textit{corrected proper momentum} $q_{i}\equiv a \, p_{i}$, $q_{j} = q \, \hat{n}_{j}$,
where $\mathbf{\hat{n}}$ is its direction unit vector.
The proper momentum $p_{i}$ is defined in terms of
the canonical conjugate momentum $P_{i}=a(1-\psi)p_{i}$ of
the the comoving coordinate $x^{i}$.
The comoving proper energy is given by $\epsilon \equiv a(p^{2}+m_{B}^{2})^{1/2}$.
Due to the perturbed geometry, we shall consider small deviations from LTE:
\begin{equation}
f(x^{i},P_{j},\tau)=f_{0}(q)\left[1+\Psi(x^{i},q,\hat{n}_{j},\tau) \right], \label{f}
\end{equation}
where $f_{0}$ is the homogenous phase-space distribution function in the ultrarelativistic limit
 \begin{equation}
f_{0}(q)=\frac{1}{e^{q/T_{B}} - 1}.
\end{equation} 
$\Psi$ is related to the temperature $T_{B}$ of
the ULBDM and its perturbation $\delta T_{B}$ in the following way:
\begin{equation}
 \Psi(x^{i}, \, q, \, \hat{n}_{j}, \, \tau)=-\frac{\partial \ln f_{0}}{\partial \ln q}(q)\frac{\delta T_{B}}{T_{B}}(x^{i}, \, q, \, \hat{n}_{j}, \, \tau).
\end{equation} 
Equation (\ref{f}) can be interpreted as a linear statistical perturbation
induced by linear metric perturbations.
In Fourier space, the Vlasov equation $df/d\tau = 0$ reads
\begin{equation}
  \dot{\Psi}-i\frac{q}{\epsilon}\Psi=-\left( (\mathbf{k} \cdot \mathbf{\hat{n}}) \dot{\psi}+i\frac{\epsilon}{q} (\mathbf{k} \cdot \mathbf{\hat{n}}) \phi \right) \frac{\partial \ln f_{0}}{\partial \ln q},
\end{equation} 
where $k$ is the wave number of the Fourier mode.
Note that the dependence on the direction vector $\mathbf{\hat{n}}$ arises only through $\mathbf{k} \cdot \mathbf{\hat{n}}$.
This last equation gives the response of the phase-space distribution function to the metric perturbations.
The natural way to proceed now is to expand the perturbation term $\Psi$ in a Legendre series:
\begin{equation}
 \Psi(x^{i}, \, q, \, \hat{n}_{j}, \, \tau) = \sum_{l=0}^{\infty} (-i)^{l}(2l+1)\Psi_{l}(\mathbf{k}, \, q, \, \tau)P_{l}(\mathbf{\hat{k}} \cdotp \mathbf{\hat{n}}),
\end{equation}
where the $P_{l}(\mathbf{\hat{k}} \cdotp \mathbf{\hat{n}})$ are the Legendre Polynomials
whose argument is the angle subtended between the wave number vector and the direction vector.
The problem turns in solving a hierarchical system of Boltzmann differential equations
for the many coefficients $\Psi_{l}$ of the Legendre expansion, called the multipole moments.
This Boltzmann hierarchy can be solved in any Boltzmann code
together with the Boltzmann equations of the rest of matter of the universe.
In the following section, we analyze the CMB spectrum
resulting from assuming different contents of CDM and ULBDM.
We carry out our numerical computations with a modified version of the public code CMBFAST.

\section{Testing with the CMB}\label{sec:TestCMB}

\begin{figure}
\epsscale{0.80}
\plotone{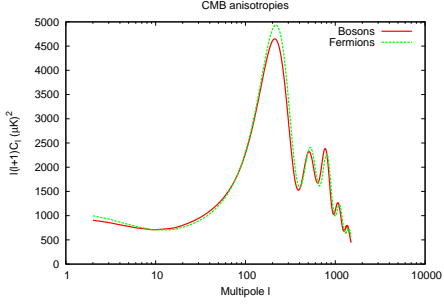}
\caption{CMB spectrums for $\Omega_{F}=\Omega_{B}=0.2$, $x_{F}=x_{B}=63109$, 
with a reionization optical depth $\tau_{r}=0.13$. All the parameters are the same for both curves.
The continuous line corresponds to bosons and the dashed line to fermions. }
\label{fig:nuVSbos}
\end{figure}

The first question is whether the CMB is effectively sensitive
to the nature of the statistics of the ULBDM.
We define the mass-to-temperature ratio of ULBDM as
\begin{equation}
x_{B} \equiv  \frac{m_{B}}{T_{B}^{(0)}},
\end{equation} 
evaluated today, with a similar definition
for the mass-to-temperature ratio of some species of fermions as $x_{F}$.
We compute the CMB spectra for fermions and bosons separetely.
All the parameters are mantained fixed in order to observe only the effect of the change of the statistics.
In Fig. \ref{fig:nuVSbos}, we show that for bosons the amplitudes of the first and second peaks are reduced;
the third peak is increased with respect to the corresponding one for fermions.
We find that for this particular set of parameters,
the response of theCMBto the change of statistics is small albeit
perceptible; we will discuss more about this behavior below.

We recall that once ULBDM decoupled,
its phase-space distribution function is freezed-out
and its temperature can only relax with the expansion of the Universe as $T_{B} \propto a^{-1}$.
After photon decoupling and in a manner quite similar to neutrinos,
the temperature of the ULBDM can be just proportional to that of photons:
\begin{equation}
 T_{B}=\alpha T_{\gamma},
\end{equation} 
 where $\alpha$ is a constant free parameter to be determined
and is only a measurement of the kinetic energy of ULBDM particles. 

\begin{figure}[b]
 \epsscale{1}
 \plotone{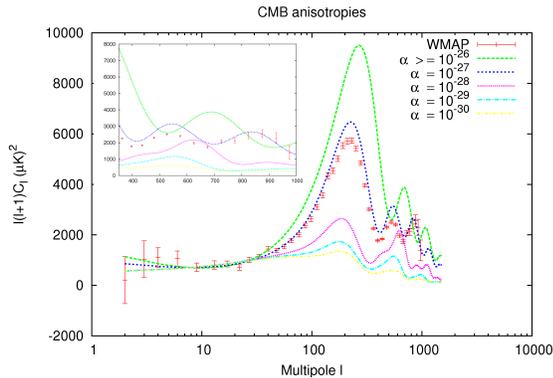}
 \caption{Response of the CMB spectrum to great changes of $\alpha$ in the interval $(\geq 10^{-26} , 10^{-30})$, with $\Omega_{B}=0.2$, $\Omega_{CDM}=0.02$. Here and in the following plots, we compare the curves to the 5yr-WMAP data of the CMB spectrum.}
 \label{fig:alfas}
\end{figure}

The radiation behaviour is defined when most of the particles in the gas are ultrarelativistic;
this happens in the limit $m_{B}/T_{B} \ll 1$ and $\omega_{B} = 1/3$.
Correspondingly, the dust behaviour occurs in the limit $m_{B}/T_{B} \gg 1$ and $\omega_{B} = 0$.
Note that if the temperature is close to that of photons, $T_{B} \approx T_{\gamma}$, $x_{B} \ll 1$,
this means that ULBDM should be still ultra relativistic today.
The effect of ULBDM on the total matter background with this value of $x_{B}$
moves the entire CMB spectrum to the right and upward;
this is shown in Fig. \ref{fig:alfas}.
For all values $\alpha \geqslant 10^{-26}$ computation of the power spectrum
results in non-sensitivity to the change of $\alpha$.
It is found then that the radiation behaviour is mantained for $x_{B} \lesssim 10^{4}$.
In all the following figures the crosses form the curve of the mean value of the observed CMB spectrum.

\begin{figure}
 \epsscale{0.80}
 \plotone{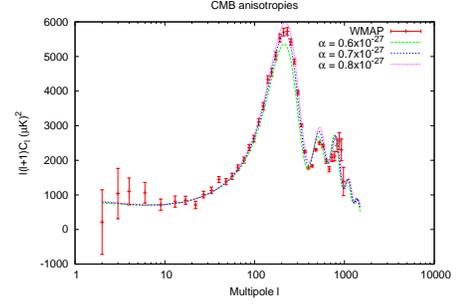}
 \caption{Response of the CMB spectrum to little changes of $\alpha$ in the interval $(0.6\times 10^{-27} , 0.8\times 10^{-27})$. For the three curves $\Omega_{B}=0.2$, $\Omega_{CDM}=0.02$.}
 \label{fig:Xs}
\end{figure}

Also in Fig. \ref{fig:alfas}, we show the prediction for $\alpha=10^{-28}$ (third curve).
In this case, ULBDM becomes non-relativistic very early,
causing a damping of the acoustic oscillations because of an increase in gravitational potential wells.
At the bottom of the same figure also appears the curve for $\alpha=10^{-30}$, in which the same effect is enhanced.
The plot shows that the order of magnitude needed to fit the data is $\alpha \sim 10^{-27}$;
this means that the mass of ULBDM must be five orders of magnitude
greater than its temperature at the present epoch ($x_{B} \sim 10^{5}$).
Of course, this is a rough estimation of $\alpha$
appropriate for the case in which ULBDM is the dominant component of DM today.
The sensitivity of the CMB power spectrum to small changes of this $\alpha$ is shown in Fig. \ref{fig:Xs}.
The range shown is from $\alpha = 0.6 \times 10^{-27}$ to $\alpha = 0.8 \times 10^{-27}$ (from $x_{B} = 73,367$ to $x_{B} = 40,759$). It is noted that the first and second peaks are enhanced if the ULBDM is more relativistic.

The above rough constraint on $\alpha$ depends of course on the relative fractions of ULBDM and CDM.
Nevertheless, quite different values of $\Omega_{B}$ and $\Omega_{CDM}$
would modify $\alpha$ less than one order of magnitude.
We then use a value of $\alpha$ pertinent for ULBDM to be dominant.

\begin{figure}[b]
 \epsscale{1}
 \plotone{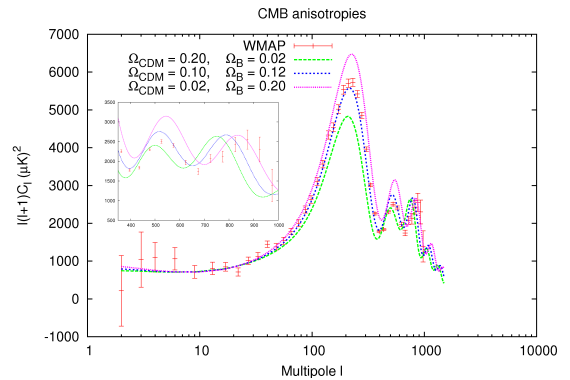}
  \caption{CMB power spectrums for different contents of ULBDM and CDM. For the three curves shown $\alpha \sim 10^{-27}$.}
 \label{fig:omegas}
\end{figure}

We now investigate the content of ULBDM against the content of CDM in two limiting cases.
One is DM dominated by ULBDM with an adequately low content of CDM and the other is the opposite case.
Fig. \ref{fig:omegas} shows how the increase in CDM diminishes the amplitude of oscillations.
This is an effect of an enhancement of gravitational potential wells of non-relativistic matter.

\begin{figure}
 \epsscale{0.80}
 \plotone{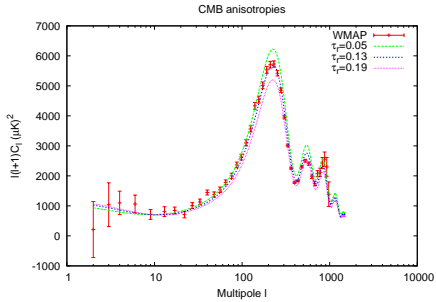}
 \caption{CMB power spectrums for different optical depths of reionization. The range shown is from $\tau_{r} = 0.05$ to $\tau_{r} = 0.19$. $\Omega_{B}=0.2$, $\Omega_{CDM}=0.02$, with $\alpha \sim 10^{-27}$.}
 \label{fig:reion}
\end{figure}

We must also consider the effect of the fraction of reionized baryonic matter.
We fix all the parameters and then we vary the reionization optical depth $\tau_{r}$,
we show in Fig. \ref{fig:reion} the values between $0.05$ and $ 0.19$.
It is found that a Universe ULBDM-dominated ($\Omega_{B}=0.2$)
seems to be allowed for $x_{B} \sim 10^{5}$ and $\tau_{r}$ about $0.07$ - $ 0.14$.
The constraints of $\tau_{r}$ from 5yr-WMAP data for $\Lambda$-CDM parameters shows a range between $0.05$ - $0.15$ ($95$\%) \citep{dunkley}.
We thus find that the mean value of our prediction $\tau_{r}=0.13$
is well within the range of the standard prediction.

Let us now return to the curves shown in Fig \ref{fig:nuVSbos}.
Note again the reduction of the first and second peaks plus the increase
of the third peak in the CMB spectrum of the bosons compared to that of fermions.
Notice that this effect is quite similar to the increase of non-relativistic matter.
Though small, it is a clear manifestation of the Pauli exclusion principle.
As is known, the energy density of Bose particles is greater than that of \textit{Fermi} particles.
This thus yields to an additional effective damping force on the acoustic oscillations,
analogous to the gravitational potential wells of the non-relativistic DM.

We now turn to interprete the thermodynamic variables.
We have found that the low mass-to-temperature ratio $x_{B}$
necessary to be the dominant component of DM today
enables us to take the the non-relativistic relationship:
\begin{equation}
 m_{B} n_{B}=\Omega_{B} \rho_{c},
\end{equation} 
where $\rho_{c}$ is the critical density of the universe.
As we have fixed the mass of ULBDM ($m_{B}=10^{-22}$ eV),
the content of ULBDM $\Omega_{B}$ determines its number density $n_{B}$.
For $\Omega_{B} = 0.2$ it follows that $n_{B} \sim 10^{25}$ cm$^{-3}$.
Such a large density would seem odd in the case of fermions (for example neutrinos)
because of the Pauli exclusion principle.
Nevertheless for ULBDM, its bosonic nature does not restrict the density of particles.
With this value of $n_{B}$ and equation (\ref{tempcrit}),
the estimation of the critical temperature is $T_{c} \approx 2.15 \times 10 ^{8}$ eV.
From the value of $x_{B}\sim 10^{5}$, or equivalently $T^{(0)}_{B} \sim 10^{-27}$ eV,
we find that the condition $T_{B} < T_{c}$ is much fulfilled.
This ensures that under the conditions to conform the dominant component of DM,
ULBDM shall necessarily be found in a BEC state today.

The explicit process of Bose-Einstein condensation during the evolution of the universe
is necessary to understand the nature and behaviour of ULBDM.
This suggests that the assumption of some kind of interactions is necessary
in order to study phase transitions from a nondegenerate state to an almost completely degenerate BEC. 

It is noteworthy that our model curves bring important,
mostly qualitative, information.
In order to provide completely conclusive results respect to the the ability of ULBDM to match the data,
it is necessary to perform further quantitative analyses.

\section{Conclusions}\label{sec:Conclusiones}

A Universe dominated by ULBDM with mass $\sim 10^{-22}$ eV could be possible
only if today the number density is of the order of $\sim 10^{25}$ cm$^{-3}$,
which implies a critical temperature of condensation $T_{c}$ about $\sim 10^{8}$ eV.
Another condition is that the mass-to-temperature ratio $x_{B}$
should be found to be about $10^{5}$,
equivalent to a temperature of the order of $\sim 10^{-27}$ eV today.
These values indicate that under the above conditions,
ULBDM is present in a Bose-Einstein condensate state.
Then, we can conclude that ULBDM endowed with an appropiate BEC
could mimic the effects of the standard CDM model on the CMB spectrum.

This value of the temperature might be falsified with more direct information
about (thermally efficient) interactions with other particles.
The energy of interaction should reveal the temperature of decoupling;
the CMB data might then provide information about coupling constants.

We have shown that changing the type of statistics in the distribution function
has non negligible effects on the CMB.
Even if not surprising,
it is interesting that the statistical nature of these two kinds of particles is perceptible in the CMB spectrum.
The effect is analogous to the addition of non-relativistic matter.

We find that the effect of reionization is necessary
to reach concordance between the ULBDM model and 5yr-WMAP data.
We do not find a substantial difference from the usual CDM prediction.
Our mean predicted value $\tau_{r} = 0.13$ is well inside the standard prediction ($95$\%).

This work might be extensible to other massive Bose gases
by means of the value of the relativistic degrees of freedom $g_{s}$,
of the particle (for scalars $g_{s}=1$, massive photons $g_{s}=3$, etc.);
of course, interactions should make the picture entirely different.
However, we restrict our discussion only to scalars in this paper
because a plausible intrinsic nature between SFDM
and cosmological BECs is found in the literature.

The next natural question is how the process of BEC formation should happen,
specifically the phase transition from a relativistic, nondegenerate gas
to a coherent classical state on the cosmological scales.
This process is expected to imply non-trivial interactions
before decoupling in the radiation epoch \citep{gabriela}.
However, this is out of the scope of the present paper and for that reason it is left for a future work.

We finally mention that the present work is an initial analysis
where we have explored only the response of the CMB to ULBDM.
It is necessary to implement a precise quantitative analysis
to the fits of all the parameters involved in the model
by using independent sets of data from other observations.

\acknowledgments
We thank L. Arturo Ure\~na-L\'opez,
Omar G. Miranda,
R. L\'opez-Fern\'andez
and Roy Maartens for useful discussions.
This work was partially supported by CONACyT M\'exico
under grants 49865-F, 216536/219673, 54576-F
and by grant number I0101/131/07 C-234/07,
Instituto Avanzado de Cosmologia (IAC) collaboration.
J.~M. gratefully acknowledges the financial support
to CONACyT project CB-2006-60526.

\end{document}